# The case for studying other planetary magnetospheres and atmospheres in Heliophysics

Ian J. Cohen[1], Chris Arridge[2], Abigail Azari[3], Chris Bard[4], George Clark[1], Frank Crary[5], Shannon Curry[3], Peter Delamere[23], Ryan M. Dewey[22], Gina A. DiBraccio[4], Chuanfei Dong[19], Alexander Drozdov[6], Austin Egert[21], Rachael Filwett[7], Jasper Halekas[7], Alexa Halford[4], Andréa Hughes[4,8], Katherine Garcia-Sage[4], Matina Gkioulidou[1], Charlotte Goetz[9], Cesare Grava[10], Michael Hirsch[14], Hans Leo F. Huybrighs[11], Peter Kollmann[1], Laurent Lamy[12,13], Wen Li[14], Michael Liemohn[22], Robert Marshall[5], Adam Masters[20], R. T. James McAteer[15], Karan Molaverdikhani[16], Agnit Mukhopadhyay[22], Romina Nikoukar[1], Larry Paxton[1], Leonardo H. Regoli[1], Elias Roussos[17], Nick Schneider[5], Ali Sulaiman[18], Y. Sun[24], Jamey Szalay[19]

1. Johns Hopkins University Applied Physics Laboratory,
2. Lancaster University
3. UC Berkeley/SSL
4. NASA/GSFC
5. University of Colorado/LASP
6. University of California, Los Angeles
7. University of Iowa
8. Howard University
9. Northumbria University
10. Southwest Research Institute
11. Khalifa University
12. LESIA, Observatoire de Paris, Université PSL, CNRS, Sorbonne Université, Université Paris Cité, Meudon
13. Aix Marseille Université, CNRS, CNES, LAM
14. Boston University
15. New Mexico State University
16. Ludwig Maximilian University of Munich
17. Max Planck Institute for Solar System Research
18. University of Minnesota
19. Princeton University
20. Imperial College, London
21. Space Dynamics Laboratory
22. University of Michigan
23. University of Alaska, Fairbanks
24. Peking University

## Summary

Heliophysics is the field that "studies the nature of the Sun, and how it influences ***the very nature of space*** — and, in turn, ***the atmospheres of planetary bodies*** and the technology that exists there."[1] However, NASA's Heliophysics Division tends to limit study of planetary magnetospheres and atmospheres to only those of Earth. This leaves exploration and understanding of space plasma physics at other worlds to the purview of the Planetary Science and Astrophysics Divisions. This is detrimental to the study of space plasma physics in general since, although some cross-divisional funding opportunities do exist, vital elements of space plasma physics can be best addressed by extending the expertise of Heliophysics scientists to other stellar and planetary magnetospheres. However, the diverse worlds within the solar system provide crucial environmental conditions that are not replicated at Earth but can provide deep insight into fundamental space plasma physics processes. Studying planetary systems with Heliophysics objectives, comprehensive instrumentation, and new grant opportunities for analysis and modeling would enable a novel understanding of fundamental and universal processes of space plasma physics. As such, the Heliophysics community should be prepared to consider, prioritize, and fund dedicated Heliophysics efforts to planetary targets to specifically study space physics and aeronomy objectives.

*The Universality of Space Physics*

Heliophysics is the field that "studies the nature of the Sun, and how it influences **the very nature of space** — and, in turn, **the atmospheres of planetary bodies** and the technology that exists there"[1]. However, NASA's Heliophysics Division tends to limit study of planetary magnetospheres and atmospheres to only those of Earth. This leaves exploration and understanding of space plasma physics at other worlds to the purview of the Planetary Science and Astrophysics Divisions. This is detrimental to the study of space plasma physics in general since, although some cross-divisional funding opportunities do exist, vital elements of space plasma physics can be best addressed by extending the expertise of Heliophysics scientists to other stellar and planetary magnetospheres. This stove-piping severely limits the broader impacts of the Heliophysics community and is detrimental to the understanding of geospace. These programmatic restrictions – and frequent non-selection of comparative studies that *are* proposed - generally limit the opportunities to study the fundamental processes of greatest scientific interest to members of the Heliophysics community to relatively few targets – primarily Earth, the Sun, solar wind, and local interstellar space. These targets are increasingly not the only - and increasingly, not the most ideal - environments in which to advance knowledge of physical processes. This has effectively hampered scientific advancement by artificially restricting our ability to codify a fundamental understanding of Heliophysics as defined above.

**The diverse worlds within the solar system provide crucial environmental conditions that are not replicated at Earth but can provide deep insight into fundamental space plasma physics processes.** One of the best ways to learn about our world and where it resides on the planetary spectrum is to study the diversity of magnetospheric and atmospheric systems and processes that exist on our neighboring worlds. The diverse planetary systems within our solar system and beyond provide crucial data points that can provide deep insight into the fundamental physics that govern our local Heliophysics environment. **Before we can claim to have a complete physics-based understanding of magnetosphere-atmosphere interactions with the Sun, we first need to test our understanding in nearby planetary systems beyond Earth.**

Shared interest in other planetary targets was already emphasized in the 2013 Solar and Space Physics Decadal Report, which stated: *"The magnetospheres of other planets display not only certain close similarities, such as the formation of bow shocks and radiation belts but also many processes that are markedly different, such as the source of charged particles within the radiation belts...Both planetary and magnetospheric understanding is thus enriched by the comparative study of magnetospheres."*[2] Of course, past collaborations with the NASA Planetary Science Division have been extremely fruitful. Some examples of these collaborations include both the shared pioneering use of the Voyager spacecraft at the outer planets and then the outer heliosphere as well as the Cassini/INCA acquisition of novel high-energy energetic neutral atom (ENA) images of the heliospheric boundary. The MAVEN mission has revolutionized our understanding of the atmospheric loss of Mars and is already informing questions about these processes at Earth. Other Planetary Science missions have also obtained serendipitous space physics measurements when possible, such as sampling of the distant solar wind, pickup ions, suprathermal particles, and cosmic rays by New Horizons and observations of interplanetary ultraviolet emissions by MESSENGER. Likewise, Heliophysics missions like Ulysses have made measurements relevant to planetary scientists, including in-situ sampling of cometary tails and the Jovian magnetosphere during its gravity assist. Such collaborations persist today, with Heliophysics-focused missions like Parker Solar Probe and Solar Orbiter making measurements during flybys of Venus. Likewise, the BepiColombo mission has been contributing to heliophysics science by consistently collecting solar wind observations during its interplanetary cruise to Mercury. Both Cassini and Juno include impressive space physics instrumentation and have thus contributed significantly to our understanding of the magnetospheres of the Gas Giants.

The Planetary Science community also highlighted the importance of comparative planetology: "Comparative planetary studies offer great potential to improve our understanding of planetary systems in general…Understanding the interactions of the solar wind at all of the planets aids in understanding the physical processes at Earth."[3] Despite this recognition, space plasma instrumentation is increasingly an afterthought on Planetary Science missions, if it is included at all. For example, the high-energy particle instruments were descoped from the Europa Clipper payload early in its formulation. The planned plasma measurements will be coarse and are only meant as a tool to support characterizing the magnetic induction from Europa's subsurface ocean. Likewise, the recent Uranus Orbiter and Probe concept - the highest priority, new large-scale mission in the recent Planetary Science and Astrobiology Decadal Study - allocates only 11.8 kg and 11 W for its entire "fields and particles package" (compared to >100 kg for in-situ instruments on Cassini). Furthermore, neither of the Discovery missions planned for Venus (i.e., VERITAS and DAVINCI+) will carry any instruments of significance for space physics research, as has been the general case for the large majority of past Discovery missions.

**Truly unlocking the mysteries of planetary magnetospheres and atmospheres throughout the solar system requires more dedicated investigations. Studying planetary systems with Heliophysics objectives, comprehensive instrumentation, and new opportunities for analysis and modeling would enable a novel understanding of space plasma physics.** Such an improved, more comprehensive understanding will also enable advancement in understanding more remote astrophysical systems, including exoplanets. A trend in this direction has already begun. Heliophysics is funding the ESCAPADE mission to Mars and has taken over control of the Radiation Assessment Detector (RAD) experiment on the Curiosity rover. The National Science Foundation (NSF) has encouraged comparative magnetospheric studies in its most recent call for the Geospace Environment Modeling (GEM) program. NASA Heliophysics has also funded mission concept studies to explore the local interstellar medium, the radiation belts of Jupiter, and the magnetosphere of Uranus. *However, as previously mentioned the Heliophysics community cannot assume that Planetary Science missions will make sufficient space physics-focused measurements.* The Heliophysics community should certainly continue to expand and enable cross-Divisional opportunities to achieve interdisciplinary science[4], with the understanding that such collaborative efforts are essential to breaking down barriers in our scientific understanding. However, *the Heliophysics Division should also be prepared to consider, prioritize, and fund dedicated Heliophysics efforts to planetary targets to specifically study space physics and aeronomy objectives.* In doing so, the community can use this new expansive ambition and broader purview to leverage additional funding so that these new endeavors will not come at the expense of continuing more historic exploration of the Sun, solar wind, and near-Earth space.

*A Solar System Full of Compelling Targets*

*Looking into the future, Heliophysics should*: proactively explore the induced magnetosphere of Venus to determine how small rocky bodies without internal magnetic fields interact with their host star; fly into the heart of Jupiter's intense radiation belts to understand the fundamental acceleration processes at play there; explore the complex and unusual magnetospheric configurations and dynamics of the Ice Giants, Uranus and Neptune; and compare and contrast other magnetospheres to Earth's and extrapolate to other, more extreme astrophysical and exoplanetary magnetospheres. These needed investigations can be achieved either by bold new standalone Heliophysics-dedicated missions to other planets or through greater collaboration with other Divisions or by bold new standalone Heliophysics-dedicated missions to other planets.

Throughout the solar system, we can combine the ground truth of in-situ particle measurements with simultaneous remote measurements equivalent to what is done for extrasolar objects. The space environments of the worlds in our solar system offer our best opportunity to directly access a natural laboratory where we can study processes that occur throughout the universe. We can learn so much more about the fundamental physical processes in the universe by adding in-situ measurements from additional data points to those of Earth, specifically those that may be more analogous to other astrophysical systems (i.e., with relativistic particle acceleration, very strong and rapidly rotating magnetic fields, synchrotron electromagnetic emissions, natural X-ray sources, etc.).

The solar system is ripe with compelling space physics targets. Below are some areas - by no means a comprehensive list - of general space physics interest that can be uniquely investigated by targets across the solar system. In many areas, the information and conditions afforded by these other bodies provide phenomena and dynamics that are beyond what can be found at Earth and thus further our understanding of our planet and the range of possibilities beyond it.

*Solar Wind-Magnetosphere Interactions*

The terrestrial magnetosphere is a solar wind-driven system dominated by the Dungey cycle; it is often used as the archetype for a solar wind-driven system in contrast to that of Jupiter, which is largely believed to be driven by internal mechanisms. However, Mercury boasts the unique combination of a weak internal magnetic field and close proximity to the Sun, producing an Earth-like magnetosphere which is possibly the most solar-wind-driven in the heliosphere. Mercury's magnetospheric response to dynamic changes in the solar wind in poorly understood and is likely a more apt archetype of a solar-wind driven system with a planetary magnetic field than Earth's respective interaction. Venus' interaction with the solar wind-sourced interplanetary magnetic field drives currents within its ionosphere and potentially metallic core, which create an induced magnetosphere including an extended magnetotail. Mars presents yet another distinct magnetospheric category in the inner planets, as described above as a hybrid magnetosphere with magnetization only due to strong located magnetic field patches in the crust across the surface of the planet which extend thousands of km in altitude. Mars' interaction with the solar wind largely occurs via currents that link to the ionosphere; its small, incredibly strong localized patches of surface magnetization in its crust create areas where local magnetic fields block the access of the solar wind to the ionosphere. The complexity of interactions with the solar wind that exist within our own solar system should be leveraged to increase our understanding of our own planet and of the fundamental interactions that are shared between planets.

In the outer solar system, the Uranian magnetosphere presents one of the most unique and underexplored data points in the solar system. With the planetary rotation axis tilted by 98° relative to the ecliptic and a highly-tilted magnetic field axis (~59°), the orientation of the magnetic field presents an asymmetrical obstacle to the impinging solar wind that varies dramatically on diurnal and seasonal timescales. Similarly, Neptune's magnetic field varies from "open" (i.e., dipole approximately aligned with the solar wind flow) to "closed" (i.e., dipole approximately orthogonal to the solar wind flow) every Neptunian day. The parameter space that these systems provide observations of cannot be observed at Earth, but can help us understand its magnetosphere (both today and in the past) as well as those of other planetary bodies.

Other objects such as asteroids and comets, can also be responsible for large magnetospheres in the solar system. For example, comets can develop plasma tails of length scales of multiple AU. As they journey through the solar system, their often highly elliptical orbits mean that they are exposed to a wide variety of solar wind and insolation conditions. When bodies near the Sun, the ices on the surface sublimate to create a dense atmosphere that expands unhindered into space and interacts with the solar wind creating a variety of magnetospheric phenomena like bow shocks and contact surfaces that are often observed at other solar system bodies, but at very different length scales. Therefore, cometary plasma environments are ideal laboratories to study cross-scale plasma effects.

*Plasma-surface interactions*

Interactions between the space environment and the surfaces of planetary bodies - specifically airless bodies such as many planetary moons and other small bodies - are extremely common in the solar system, but are not well understood. For example, we still do not understand what balance of surface processes releases neutral atoms and molecules into the exosphere of Mercury and what role they play in coupling the exosphere to the magnetosphere. Likewise, it is not understood whether transient plumes from objects such as Ceres can generate a sporadic bow shock and whether some asteroids (i.e., Psyche) have strong magnetic fields capable of standing off the solar wind. Even at the Moon, questions abound as to whether small-scale magnetic fields are capable of producing plasma processes such as collisionless shocks and/or reconnection. In the outer solar system, there are significant questions about how radiation in planetary magnetospheres may affect the surface composition of icy moons around Jupiter, Saturn, and Uranus.

*Auroral Processes and Current Systems*

The terrestrial aurora remains a major focus of study of magnetosphere-ionosphere coupling, with decades of observations of the global field-aligned current system and auroral precipitation. However, auroral emissions have also been observed at several other planets in the solar system with unique characteristics. For example, despite having only regional "mini magnetospheres", multiple distinct auroral processes are still found at Mars. Neptune's magnetospheric configuration generates a unique configuration where the plasma sheet becomes cylindrical; a major mystery persists as to how such a complex current sheet would close. As with many things, Jupiter has the most intense auroral emission in the solar system, which seems to be largely decoupled from solar wind interactions and is instead dominated by internal processes. In particular, Jupiter is known for the presence of auroral footprints associated with its largest Galilean moons that offer a new window to study the interaction and dynamics of Alfven waves. New results from the Juno mission suggest that potentials in the auroral regions reach into the megavolts[5], raising the question as to why these potentials are two orders of magnitude lower at the Earth. If auroral energies are so extreme as at Jupiter they may even provide a seed mechanism for the acceleration of the energetic particles of its radiation belts.

Additionally, though the detection from exoplanets has yet to be confirmed, we have found such emissions from larger objects such as ultracool dwarfs, providing measurements of their magnetic field and insight into the presence of electrons into the MeV range[6]. The search for exo-magnetospheres at radio wavelengths with ground-based telescopes has become a hot topic. Understanding auroral radio emissions from the solar system with in-situ measurements is more than ever essential to build up the reference frame from which we will interpret distant radio emissions from magnetospheres that will never be explored via in-situ spacecraft.

*Magnetospheric Transport and Dynamics*

The transport and dynamics of plasma are defining characteristics of a planetary magnetosphere. Earth's magnetosphere is often used as a template from which we can understand other planetary environments. However, as our understanding of other magnetospheres in the solar system has grown, we have begun to reassess how common Earth's magnetosphere may truly be. Even our limited understanding of the magnetospheres in our solar system suggests that it is unlikely that the Dungey cycle is a universal magnetospheric process.

Other planetary bodies exhibit interesting transport mechanisms and can be used to compare and contrast to that of Earth. While small, Mercury's magnetosphere has most of the primary characteristics seen in planetary magnetospheres, including energetic particles and dynamic phenomena (e.g., substorms and particle injections). At Venus, planetward plasma flows have been observed downstream of the planet, the drivers of which remain a mystery given Venus' lack of a magnetic field. Saturn's magnetosphere is unique because its rotation axis and its magnetic dipole axis are co-aligned with each other. It appears to interact with the solar wind, driven by a process that is similar to the Dungey cycle; however, like Jupiter, it also has a moon that serves as the system's primary plasma source. This adds additional plasma physics interactions (e.g., neutral-charged interactions) that offer unique insights into such fundamental processes. Farther out in the solar system, limited observations from Voyager 2 make it difficult to fathom the transport mechanisms in Uranus' complex magnetospheric configuration. Because Uranus is a fast rotator and the magnetosphere changes between being open and closed to the solar wind throughout a Uranian day, planetary rotation must also play some role in driving plasma flow. Unlike other planets, Uranus' corotational electric field can seasonally become perpendicular to the convection electric field. This unique magnetospheric configuration and dynamics can serve as a prime laboratory in which to generally understand plasma flows in a magnetosphere. At other planets, it is difficult to disentangle processes such as whether flows result from tail reconnection of centrifugally-driven interchange instabilities. Last, but certainly not least, Neptune's magnetosphere is largely composed of protons and nitrogen assumed to be from its largest moon Triton, believed to be a captured Kuiper Belt object. Unlike other planetary magnetospheres in the solar system, where such internal sources build up and energize over time, Voyager 2 found Neptune's magnetosphere to have surprisingly low ion densities. It remains unclear how this magnetospheric plasma may be lost from the system, as no clear signatures of magnetic reconnection were observed. In summary, exploring and understanding the plasma transport and dynamics at play in other planetary magnetospheres throughout the solar system can help us better understand how typical Earth's magnetosphere may be.

*Radiation Belts and Particle Acceleration and Loss*

Planetary radiation belts magnetically trap and energize charged particles around a planet and are as diverse as the planets they encompass. While Earth's radiation belts have been recently studied in depth by the Van Allen Probes mission, other solar system targets provide new and unique domains through which we can learn about radiation belt, acceleration, and loss processes. For example, the acceleration mechanism of the few-keV planetary ions found at Mercury and what role they play in magnetospheric dynamics remain unknown. Likewise, the Jovian system has the most extreme radiation belts in the solar system, including high-intensity populations of several tens of MeV electrons. Discovering the nature of such extremely efficient acceleration of relativistic electrons will greatly advance our understanding of processes that drive particle acceleration during rare terrestrial space weather events that make them difficult to study with good spatial coverage, as well as acceleration throughout the universe, such as at Giant planets and possible exoplanetary magnetospheres, and the production of cosmic rays, synchrotron, and x-ray emissions in astrophysical objects and events. Further out, Uranus challenges our understanding of radiation belt physics because it has electron radiation belts that are similar in intensity up to MeV energies as those of Earth and Jupiter, despite having a sparse magnetospheric source population of low energy plasma, slow acceleration through radial diffusion, and the strongest whistler-mode hiss and chorus waves observed by the Voyager spacecraft[7]. Finally, it's unclear how the presence of a captured Kuiper Belt Object - i.e., Triton - affects the evolution and dynamics of the radiation belts of Neptune.

*Planetary Atmospheres and Aeronomy*

The planetary atmospheres across the solar system provide a huge range of diversity and opportunities for comparative study. Nearly a dozen different escape processes have been identified in our solar system, and understanding how the dominant mechanisms vary at different planets and over time is important in understanding the evolution of atmospheres and conditions necessary for habitability. For example, Venus offers an ideal counterpart to Earth and Mars when studying atmospheric evolution: it is thought that the relatively direct access of the solar wind to Venus' atmosphere was responsible for the removal of water from its atmosphere, while it remained at Earth and Mars. The fast rotation speeds of the Giant planets drive atmospheric dynamics very different from those found on Earth. For example, the global atmospheric circulation patterns at the Ice Giants remain largely unknown as do the effects of their complex magnetic topologies on their magnetosphere-ionosphere coupling dynamics and atmospheric composition. At Jupiter and Saturn, significant wind shears drive complex atmospheric dynamics and intense storms such as the Great Red Spot. Furthermore, Titan's dense atmosphere and Triton's intense ionosphere raise many questions about moon aeronomy.

Furthermore, asteroids and comets far away from the Sun have barely any atmosphere, so that sputtering and surface weathering produce a thin ionosphere under low gravity conditions, thus providing a unique opportunity to study a rapidly-expanding atmosphere.

*Moon-magnetosphere interactions*

Unlike the Earth, the outer planets all have moons which likely play significant roles in shaping their magnetospheric dynamics. Saturn's magnetosphere is rotationally-powered (as opposed to solar wind-driven) and has a major provider of neutrals and plasma (its moon Enceladus) embedded deep within the system. As a result, neutrals dominate the magnetospheric environment, with more than one hundred times more neutrals than ions, so that the equatorial neutral density is usually significantly larger than the plasma density. Conversely, plasma introduced from Io dominates the Jovian system and likely provides a source for its intense radiation belts; the strongest whistler-mode waves in Jovian magnetosphere are observed in vicinity of Ganymede and Io. At the solar system's edge, the role of Triton - an expected active ocean world - in the dynamics of Neptune's magnetosphere persists as an outstanding question.

Moons remove particles from the surrounding magnetosphere, by absorption, deflection, and/or scattering. This leads to extended wakes with low intensities of energetic particles. These well-defined signatures are ideal tools to track magnetospheric dynamics in a way that is not possible at the Earth. The gradual refilling of the absorption is an unambiguous measure of radial diffusion[8]. The displacement of the absorption feature allows tracking radial plasma flows that occurred before the particles reached the spacecraft and can track the dynamics of convection[9].

*Fundamental Plasma Processes*

The variation in the characteristics of the solar wind and interplanetary magnetic versus radial distance from the Sun can lead to significant differences in the plasma environments surrounding the planets - and thus also the conditions present to potentially enable fundamental plasma processes such as magnetic reconnection and collisionless shocks.

Magnetic reconnection is known to play a significant role in the transport of mass, energy, and momentum into and throughout the terrestrial magnetosphere. Typical inner solar system solar wind characteristics result in a reconnection rate at Mercury's dayside magnetopause that is approximately ten times that found at Earth. It is not understood what role kinetic effects play in phenomena like Kelvin-Helmholtz boundary waves and magnetopause reconnection. There have also been magnetic fields, plasma flows, and electron signatures suggestive of magnetic reconnection observed at Venus. Conversely, we do not know whether magnetic reconnection plays an important role in the global magnetospheric dynamics at outer planets.

The bow shocks of the outer solar system planets exhibit much higher Mach numbers than available in the inner solar system. Saturn's bow shock, for example, was shown to occasionally attain Mach numbers >100 and evidence has been found that parallel shock acceleration is much more important than predicted[10]. Farther out in the heliosphere, the likelihood of detecting strong shocks becomes greater.

*Conclusion*

In summary, the diverse worlds within the solar system provide crucial environmental conditions that are not replicated at Earth but can provide deep insight into fundamental space plasma physics processes. Studying planetary systems with Heliophysics objectives, comprehensive instrumentation, and new grant opportunities for analysis and modeling would enable a novel understanding of fundamental and universal processes of space plasma physics. As such, the Heliophysics community should be prepared to consider, prioritize, and fund dedicated Heliophysics efforts to planetary targets to specifically study space physics and aeronomy objectives.